\documentclass{article}

\usepackage{PRIMEarxiv}

\usepackage[utf8]{inputenc} % allow utf-8 input
\usepackage[T1]{fontenc}    % use 8-bit T1 fonts
\usepackage{hyperref}       % hyperlinks
\usepackage{url}            % simple URL typesetting
\usepackage{booktabs}       % professional-quality tables
\usepackage{amsmath,amssymb,amsfonts}       % blackboard math symbols
\usepackage{nicefrac}       % compact symbols for 1/2, etc.
\usepackage{microtype}      % microtypography
\usepackage{lipsum}
\usepackage{fancyhdr}       % header
\usepackage{graphicx}       % graphics
\graphicspath{{media/}}     % organize your images and other figures under media/ folder
\usepackage{float}
\usepackage{algorithmic}

\title{Towards the "Digital Me": A vision of authentic Conversational Agents powered by personal Human Digital Twins}

\author{
  Lluís C. Coll \\
  Technical University of Denmark \\ 2800 Kgs. Lyngby \\ Denmark \\
   \And
  Martin W. Lauer-Schmaltz \\
  Technical University of Denmark \\ 2800 Kgs. Lyngby \\ Denmark \\
   \AND
  Philip Cash \\
  School of Design \\ Northumbria University \\ Newcastle upon Tyne NE1 8ST \\ United Kingdom \\
   \AND
   John P. Hansen \\
  Technical University of Denmark \\ 2800 Kgs. Lyngby \\ Denmark \\
  \AND
  Anja Maier \\
  Department of Design, Manufacturing and Engineering Management \\ University of Strathclyde \\ James Weir Building, Office JW701a \\ 75 Montrose Street \\ Glasgow G1 1XJ \\ United Kingdom \\
}

\begin{document}
\maketitle

\begin{abstract}
Human Digital Twins (HDTs) have traditionally been conceptualized as data-driven models designed to support decision-making across various domains. However, recent advancements in conversational AI open new possibilities for HDTs to function as authentic, interactive digital counterparts of individuals. This paper introduces a novel HDT system architecture that integrates large language models with dynamically updated personal data, enabling it to mirror an individual's conversational style, memories, and behaviors. To achieve this, our approach implements context-aware memory retrieval, neural plasticity-inspired consolidation, and adaptive learning mechanisms, creating a more natural and evolving digital persona. The resulting system does not only replicate an individual's unique conversational style depending on who they are speaking with, but also enriches responses with dynamically captured personal experiences, opinions, and memories. While this marks a significant step toward developing authentic virtual counterparts, it also raises critical ethical concerns regarding privacy, accountability, and the long-term implications of persistent digital identities. This study contributes to the field of HDTs by describing our novel system architecture, demonstrating its capabilities, and discussing future directions and emerging challenges to ensure the responsible and ethical development of HDTs.
\end{abstract}

% keywords can be removed

\section{Introduction}
In recent years, Human Digital Twins (HDTs) have emerged as a transformative technology with the potential to revolutionize applications in various sectors from manufacturing to healthcare. HDTs are comprehensive, continuously evolving digital representations of human individuals that mirror their physiological, cognitive, and behavioral characteristics \cite{lauerschmaltz2024}. Their ability to monitor, simulate, and predict human conditions has shown promising potential, for instance, in facilitating informed, personalized clinical interventions and enhancing ergonomic safety and efficiency in industrial settings through individualized feedback \cite{Okegbile2023,Wang2024}. However, while most authors to date view HDTs primarily as evolving databases enhanced with AI and simulation capabilities for personalized feedback and decision support, one might question whether future HDTs should remain as mere data models or further evolve into more sophisticated and lively virtual counterparts of human individuals. For instance, some authors suggest that HDTs could further incorporate deep human cognitive mechanisms, personal knowledge, behavioral and personality characteristics, and autonomy and agency, thus indicating the so-far untapped potential of HDTs beyond their current role as systemic utilities \cite{Saariluoma2021, Lin2024}. This raises the question of what alternative conceptualizations and applications HDTs could embrace to open up new possibilities across various domains. \par
One such alternative conceptualization of HDTs is the vision of a holistic virtual co-existence of oneself—like a virtual twin sibling that looks, sounds, talks, and behaves like the individual it represents. With the rapid advancement of Large-Language Models (LLM) and other AI technologies, the notion of creating such digital replicas of humans as natural, interactive virtual companions has gained momentum, giving birth to a variety of applications, ranging from personal assistant systems, such as Google, Alexa, Siri, and ChatGPT, to platforms offering virtual friends and partners \cite{Strohmann2023}. Furthermore, some platforms even allow individuals, such as celebrities, to create virtual versions of themselves, allowing their fans to interact and become friends with these virtual personas \cite{tolentino2023snapchat, vendrell2023celebrities, zhang2023online}.\par 
These initiatives not only present considerable commercial opportunities (see Section \ref{subsubsec:Discussion:VisionsAndFutureApplications}) but also hold the potential to enhance individuals’ lives, for instance, by facilitating communication for individuals facing communicative challenges, such as stroke survivors with Aphasia, unable to express their needs and emotions themselves \cite{Fotiadou2014}. However, while such applications would require the virtual representation to integrate sophisticated knowledge about the represented individual, recent representational agents and avatar systems primarily focus on replicating their human counterparts' external attributes and communication styles \cite{aneja2019,Hoegen2019,Ou2023}. Thus, to attain a truly comprehensive digital representation, future developments must additionally establish a continuous, dynamic link that allows for a co-evolving relationship between the digital “twin” and its human counterpart to obtain a deeper understanding of an individual's unique characteristics, preferences, and memories.\par 
Addressing this challenge, this paper introduces a novel system architecture that effectively leverages the conversational and cognitive capabilities of GPT-4o alongside the rich personal knowledge base and dynamic updating mechanisms inherent in HDT technology to reinterpret the concept of HDTs as autonomous, believable virtual counterparts of specific human individuals. Our resulting system fuses general personal information with dynamic data streams, such as conversation logs as well as vital signs and activity data from wearable devices to create a virtual agent that mirrors the human individual's unique conversational style and characteristics, able to autonomously engage in conversations on behalf of the individual in an authentic and credible manner. In doing so, we address the following research question:\par
\textbf{RQ: }\textit{How can the integration of advanced conversational AI with dynamic, multimodal personal data enhance the authenticity and effectiveness of virtual representations in mimicking human personalities in interactions?}\par

By answering this research question, this paper makes several important contributions to the fields of HDTs, digital human modeling, and interactive AI systems:

\begin{itemize}
    \item It introduces a novel perspective on HDTs, redefining their potential beyond traditional roles of data support and system feedback to act as dynamic, virtual counterparts that mirror individuals’ unique conversational styles and personal characteristics.
    \item It demonstrates a novel prompt-based system architecture that combines real-time personal data with advanced conversational AI, enhancing the ability of HDTs to authentically replicate the user’s personality for more personalized and effective interactions.
    \item It explores potential future applications and reflects on ethical considerations of advanced virtual representations, laying the foundation for responsible innovation in highly personalized, interactive HDTs.
\end{itemize}

\section{Theory and Related Work}
Before delving into the details of the proposed system architecture, we first establish a basic understanding of the concept of HDTs and the recent advances in personalized Conversational Agents (CAs), along with the fundamental functionalities of the human memory system, on which we will later ground our HDT's cognitive capabilities.
\subsection{Going beyond traditional Human Digital Twins}
\label{subsec:HDT-Theory}
\begin{figure}[H]
    \centering
    \includegraphics[width=1.0\linewidth]{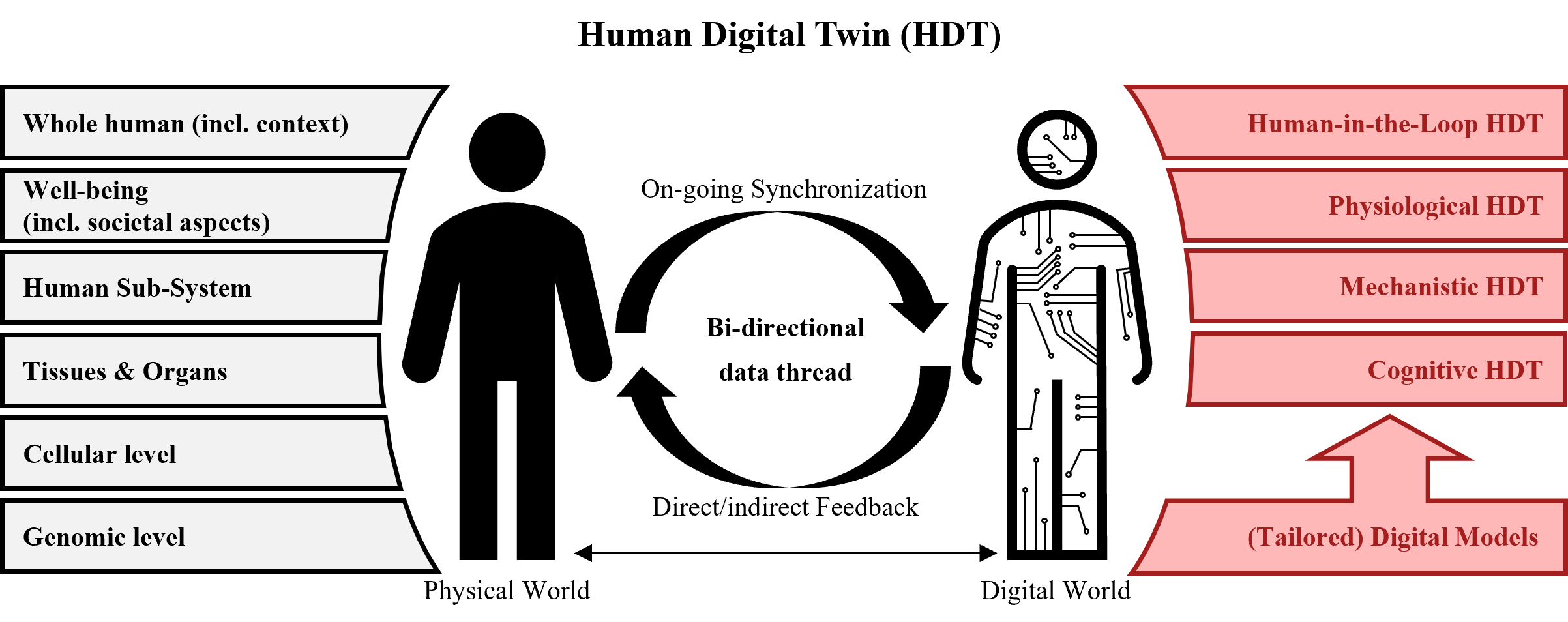}
    \caption{A general conceptualization of Human Digital Twins defined as: A digital representation of a human (or specific aspect of a human); which uses a bi directional data thread for on-going synchronization with the human’s state; and provision of feedback based on data aggregation and predictions to directly (or indirectly) influence the human entity, user(s) and/or their environment. \cite{lauerschmaltz2024}}
    \label{fig:HDT-Overview}
\end{figure}

The concept of HDTs evolved from that of traditional Digital Twins known from various domains, such as manufacturing and automotive industry. DTs, first introduced by Michael Grieves in 2003 and later named by NASA, describe dynamic, continuously updated virtual models of physical systems or processes that simulate the physical system's behavior in real-time, allowing for performance optimization, predictive maintenance, and insights into system operations through data analysis and monitoring \cite{Barricelli2019, Fuller2020}. With the realization of DTs' potential and the increasing importance of human-centric systems, this concept was eventually expanded to also encompass digital representations of individuals. The resulting HDTs today define as \textit{"Comprehensive digital representations of humans, based on multi-dimensional information regularly updated via sensor data or manual input, which provide analytical and simulated feedback relevant to the HDT’s application context."} \cite{lauerschmaltzETHICA2024}. As such, similar to traditional DTs, HDTs essentially consist of three core components: a physical entity (i.e., the human to be represented), a digital entity (i.e., the corresponding digital representation of the human), and a bi-directional data link that allows the continuous synchronization of the digital entity with its physical counterpart as well as the transmission of data-driven feedback to the physical world (Fig. \ref{fig:HDT-Overview}).

Comprehensive reviews of the traditional concept of HDTs and its recent applications are provided by various recent surveys \cite{lauerschmaltz2024, lauerschmaltzETHICA2024, Wang2024,Lin2024,SHENGLI2021,Miller2022}. Generally, HDTs leverage personal data from wearable and optical devices, medical records, social media, and manual input to construct a continuously evolving digital model of the underlying human individual and use artificial intelligence and simulation techniques to generate informed feedback and decision support for the contextual application and relevant stakeholders in various domains. For instance, in human-centered production lines, HDTs can improve ergonomic safety by monitoring and analyzing the employees' movements and optimize operational efficiency by aligning work schedules with the employees' skills and current capacity \cite{Calzavara2019,Nikolakis2019,Greco2020,FAN2022}. In contrast, HDTs in healthcare can provide deep insights into the evolution of a patient's health state to facilitate diagnostics and provide informed decision-support concerning suitable interventions and their predicted outcomes \cite{Corral2020,LAUZERAL2019,GOODWIN2020,Lal2020-ed}. Further, HDTs in smart home and personal assistance can detect behavioral patterns and preferences to customize smart devices' settings and tailor suggestions and functionalities to the human's daily routines \cite{Hafez2020,Bouchabou2023,Zibuschka2020}. Thus, HDTs to date are typically designed as dynamic data backbones for systems, aiming to inform specific context applications and their users through context-aware feedback and decision-support mechanisms.\par
In addition to this application-focused perspective, some authors have recently started to explore the potential roles of HDTs beyond that of mere systemic utilities. For instance, in our prior work, we introduced the concept of an HDT Opponent for serious game-based rehabilitation, i.e., an active in-game component blending an avatar mirroring the patient's appearance with behavior reflecting their past performances, thus transforming the HDT from a simple data backbone into an interactive element that fosters self-identification and boosts rehabilitation effectiveness by promoting self-challenge \cite{DTOpponent}. Additionally, \cite{nttHumanDigitalTwins} and \cite{microsoftDigitalMe} propose the concept of a "Digital/Another Me," envisioning HDTs as complete virtual counterparts of humans that replicate the individual's appearance, behavior, knowledge, and skills, effectively bridging the gap between the physical and virtual realms by serving as digital, autonomous representatives capable of acting on behalf of the human user in digital environments. Thus, these alternative perspectives on the concept of HDTs suggest a progressive shift towards more immersive, personalized, and autonomous digital representations with an increased focus on the individual and their seamless integration and interaction with virtual worlds.

\subsection{Advancements in personalized Conversational Agents}
\label{subsec:CA-theory}
A key aspect that defines humans and decisively shapes how others perceive us is our interpersonal interaction, particularly through language. Language--whether verbal or non-verbal--serves as a powerful medium for expressing thoughts, emotions, and identity, as well as fostering understanding and connection among individuals \cite{Firth1950}. Its unique ability to structure and convey complex information has long inspired efforts to replicate human communication in artificial systems, resulting in the rise of CAs.\par
CAs are intelligent systems designed to simulate human-like conversational abilities, aiming to enhance the naturalness and depth of interaction between humans and digital systems by authentically mirroring human conversational behavior. Having evolved significantly from their early conceptualizations as simple rule-based systems, CAs today leverage complex algorithms in the field of natural language processing and machine learning to understand and generate human-like responses, catering to a wide range of applications from customer service bots to mental health advisors \cite{Alnefaie2021,Allouch2021}. For example, \cite{Lieb2024} demonstrates how CAs powered by large LLMs can enhance education by delivering educational content on behalf of teaching staff. Additionally, LLM-based CAs have proven effective also in e-commerce, offering improved real-time customer support \cite{CHUNG2020587,Rakhra2021}. Other prominent commercial examples include virtual personal assistants, such as Siri, Alexa, and Google Assistant, able to perform tasks ranging from answering queries to controlling smart home devices, all through conversational interfaces.\par
A pivotal advancement in the field of CAs, significantly enhancing their performance and capabilities, has been the introduction of transformer models by Google \cite{vaswani2023attentionneed}. Transformer models, characterized by their ability to handle sequential data and their attention mechanisms, have set new benchmarks in NLP tasks and facilitated the development of more nuanced and context-aware LLMs, such as OpenAI's GPT-4o. This new generation of high-performance LLMs provide astonishing human-like conversation capabilities in addition to being able to remember contexts even over longer interactions, simulate empathy, and adapting their responses based on the user's preferences and past interactions. Given these pivotal capabilities, LLMs have recently been used as the backbone for cognitive and response generation modules in a large number of new CAs, including virtual assistants and companions \cite{Dong2023,LiYuanchun2024,Replika}.\par 
Additionally, some CAs are even designed to mimic specific human individuals, such as celebrities, allowing users to interact with their virtual clones. For instance, in 2023, a social media influencer with over 1.8 million followers created a virtual persona by training a GPT-based LLM on 2,000 hours of her content and combining it with a custom avatar, enabling fans to interact with the virtual version for a subscription fee per minute \cite{sternlicht2023snapchat}. However, while these systems excel in mirroring appearance, voice, and conversational style, and can adapt to user preferences over time, their knowledge about the represented individuals remains limited due to the lack real-time connection to their human counterparts, often resulting in responses that may not align with how the actual person would react in certain situations. Thus, despite the growing trend towards creating digital entities that accurately represent human individuals, the absence of ongoing knowledge exchange restricts these technologies from evolving into true HDTs as described in the previous section, underscoring the need for further development to fully bridge the physical and virtual worlds.

\subsection{Human Memory and Memory Retrieval}
\label{subsec:HumanMemoryAndRAG}
To further improve the quality of feedback and personalization of CAs, researchers and developers have recently explored ways to enrich CAs with task-relevant knowledge bases. Common approaches include the use of Retrieval-Augmented Generation (RAG) models, a technique that integrates a retrieval system with a generative language model to provide contextually relevant and accurate responses based on external knowledge sources. However, these systems are often limited by static or pre-defined knowledge repositories, which struggle to adapt to new, evolving, or highly personalized information in real-time \cite{Huang2024RAG}. They also often lack the ability to prioritize, filter, and contextualize information based on the nuances of individual users or ongoing interactions \cite{barnett2024sevenfailurepointsengineering}. Thus, authentic HDTs require more dynamic and reflective mechanisms, similar to the functionality of the human memory system, where information is not only stored but continuously updated, structured, and selectively retrieved to reflect the individual’s personality, experiences, and contextual needs.\par
\begin{figure}[H]
    \centering
    \includegraphics[width=1.0\linewidth]{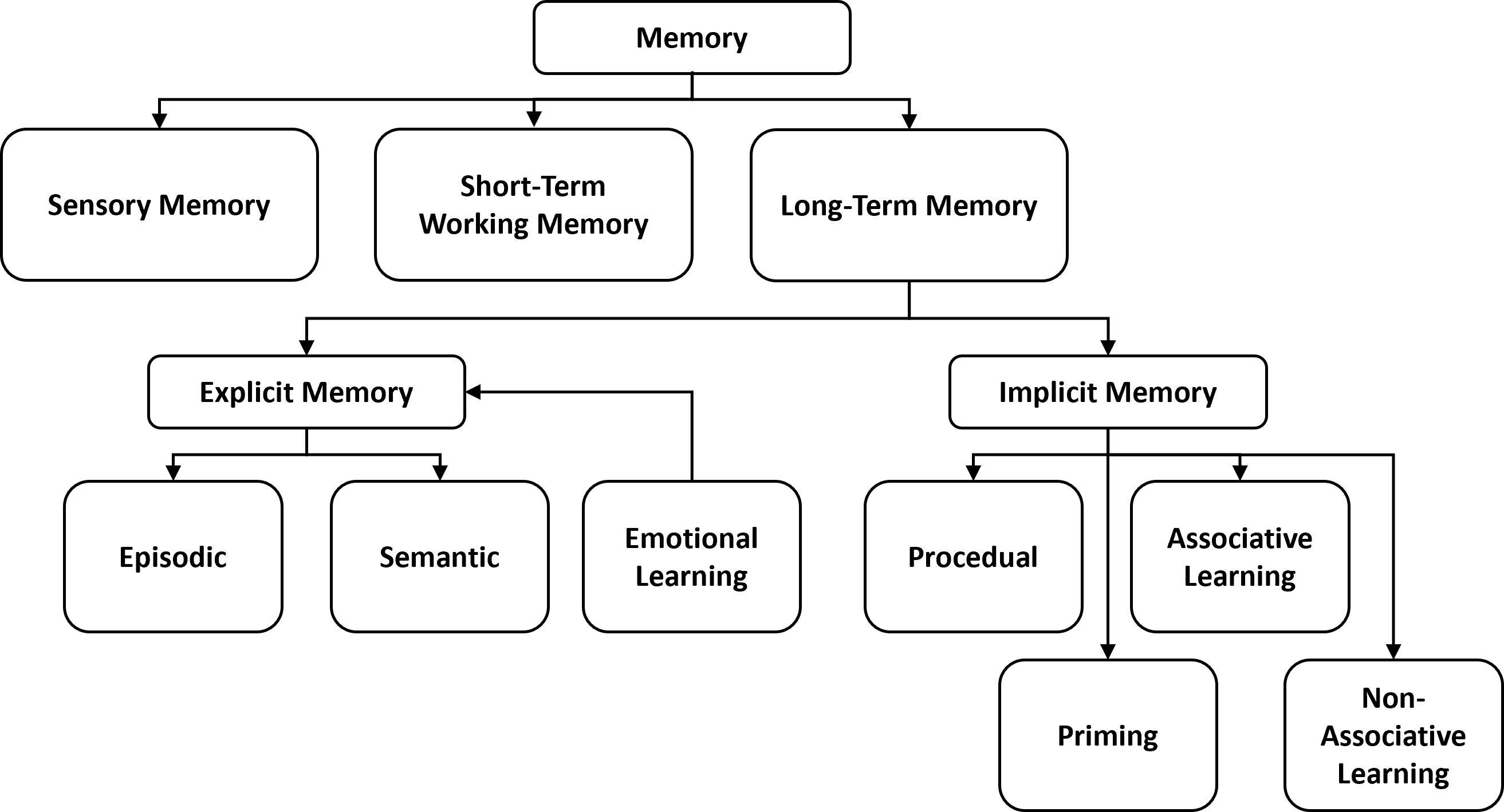}
    \caption{Simplified categorization of the types of memories within the human memory system and the brain areas they are associated with \cite{hove2024biological}.}
    \label{fig:HumanMemorySystem}
\end{figure}

The human memory system itself encompasses various types of memory (Fig. \ref{fig:HumanMemorySystem}), each serving specific purposes. While our short-term working memory temporarily holds information necessary for immediate cognitive tasks, such as remembering a phone number long enough to dial it, our long-term memory ensures the long-term preservation of information, segmented into explicit and implicit memory types. Explicit memories thereby encompass episodic memories, such as autobiographical events (e.g., the first day at school), and semantic memories, such as general knowledge and facts (e.g., what a school is and what it entails). The prevalence of these memories can further be affected by their emotional significance, making some memories, especially those with strong emotional components, more vivid and enduring. Additionally, implicit memories, operating below the level of conscious awareness, play a crucial role in our ability to perform tasks and behaviors automatically. Here, the procedual memory ensures that, once learned, we can execute specific skills, such as walking or playing instruments, without consciously thinking about them, whereas priming allows us to subconsciously derive connections from one stimulus to a subsequent stimulus that might affect our reaction to the latter. \cite{hove2024biological}\par   
Moreover, leveraging the knowledge stored within the human memory system requires a complex memory retrieval process that involves a number of interconnected, non-trivial factors. Here, it is crucial to differentiate between available information, which is all data stored in memory, and accessible information, which are the memories we can retrieve in a specific moment \cite{TULVING1966381}. The accessibility of specific memories thereby largely depends on the effectiveness of cues, with success tied to how closely these cues mirror those present during the initial encoding of information \cite{tulving1973encoding}. Additionally, retrieving memories is not a mere replication of past events or knowledge but rather a reconstructive process shaped by our current knowledge, assumptions, and the context of retrieval, making every act of memory retrieval both selective and subjective, potentially even altering the memory itself \cite{bartlett1932remembering}. Thus, the knowledge contained within the human memory system and the ability to retrieve them collectively set the cornerstone to our personality, determining our knowledge, behaviors, and skills, and enabling us to perform meaningful interactions with our environment.

\section{Towards the "Digital Me" - A novel architecture for authentic Human Digital Twins}
\label{sec:Architecture}
\begin{figure*}[!htpt]
    \centering
    \includegraphics[width=\textwidth]{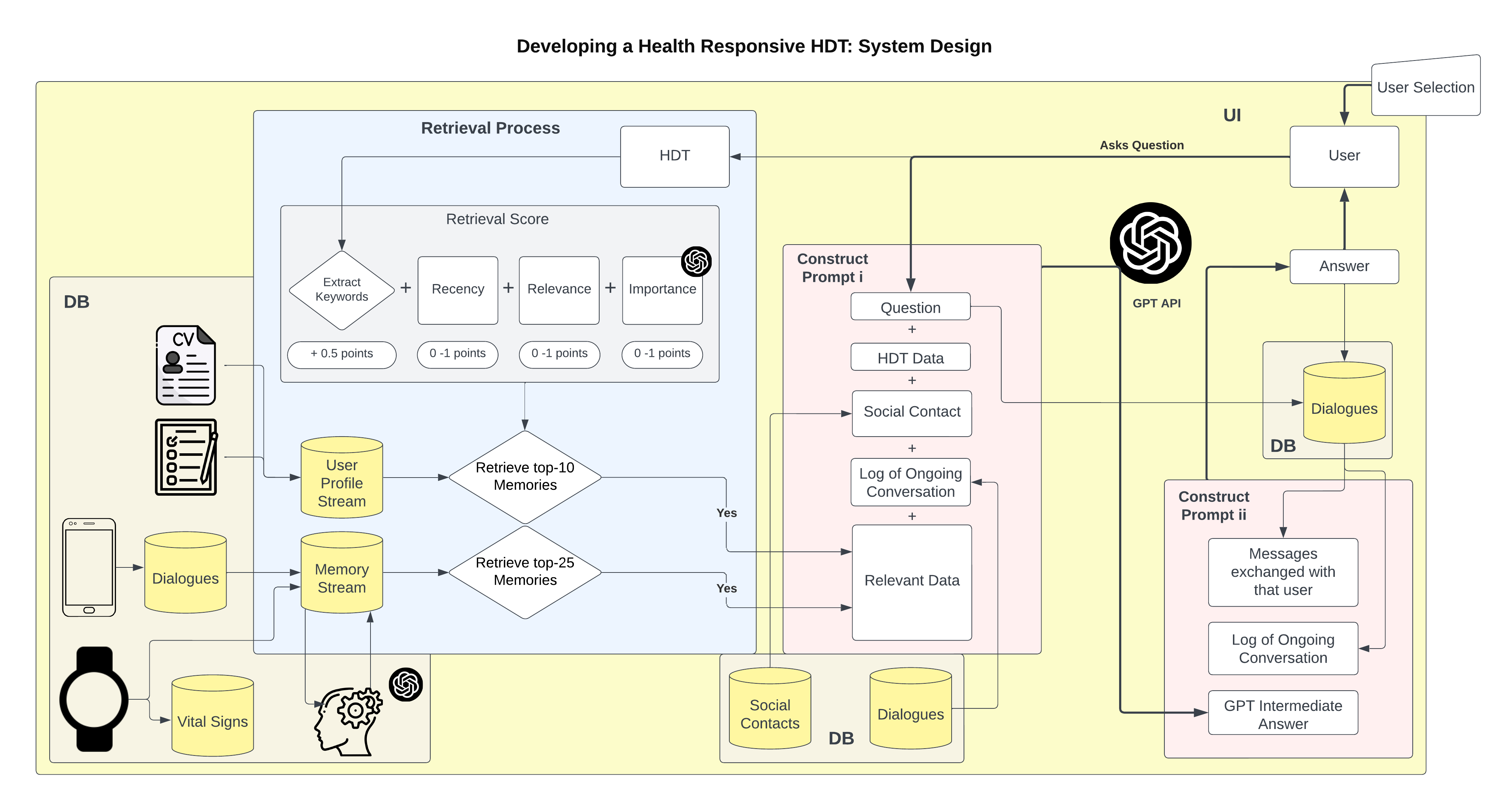}
    \caption{Architecture of the proposed HDT system, consisting of data collection and storage modules (brown), memory retrieval modules (blue), and response generation modules (pink).}
    \label{fig:HDTSystemArchitecture}
\end{figure*}

This section outlines the architectural framework and operational processes of the proposed HDT system, developed in response to our research question. Drawing on the theories of human memory described in Section \ref{subsec:HumanMemoryAndRAG}, we designed a CA system that uses decay functions to prioritize memories and leverages OpenAI's GPT-4o for in-depth reflections on conversational data and vital signs. The system enables text-based interactions that emulate the conversational style, knowledge, and memories of the person represented. Other aspects necessary for a comprehensive virtual representation, such as physical appearance and authentic vocal communication, have been extensively addressed by other researchers (e.g., \cite{Arik2018, Altundas2023}), and are beyond the scope of this paper. Instead, our focus lies on developing a system that continuously learns about the individual it represents, evolving alongside them by collecting and updating knowledge about their experiences, traits, health, and social context, while replicating their unique conversational style and content. The resulting HDT system (Fig. \ref{fig:HDTSystemArchitecture}) comprises three core components: i) data collection and storage, ii) memory retrieval, and iii) response generation.  

\subsection{Data Collection and Storage}
\label{subsec:DataCollectionAndStorage}
\begin{figure*}[!htbp]
    \centering
    \includegraphics[width=\textwidth]{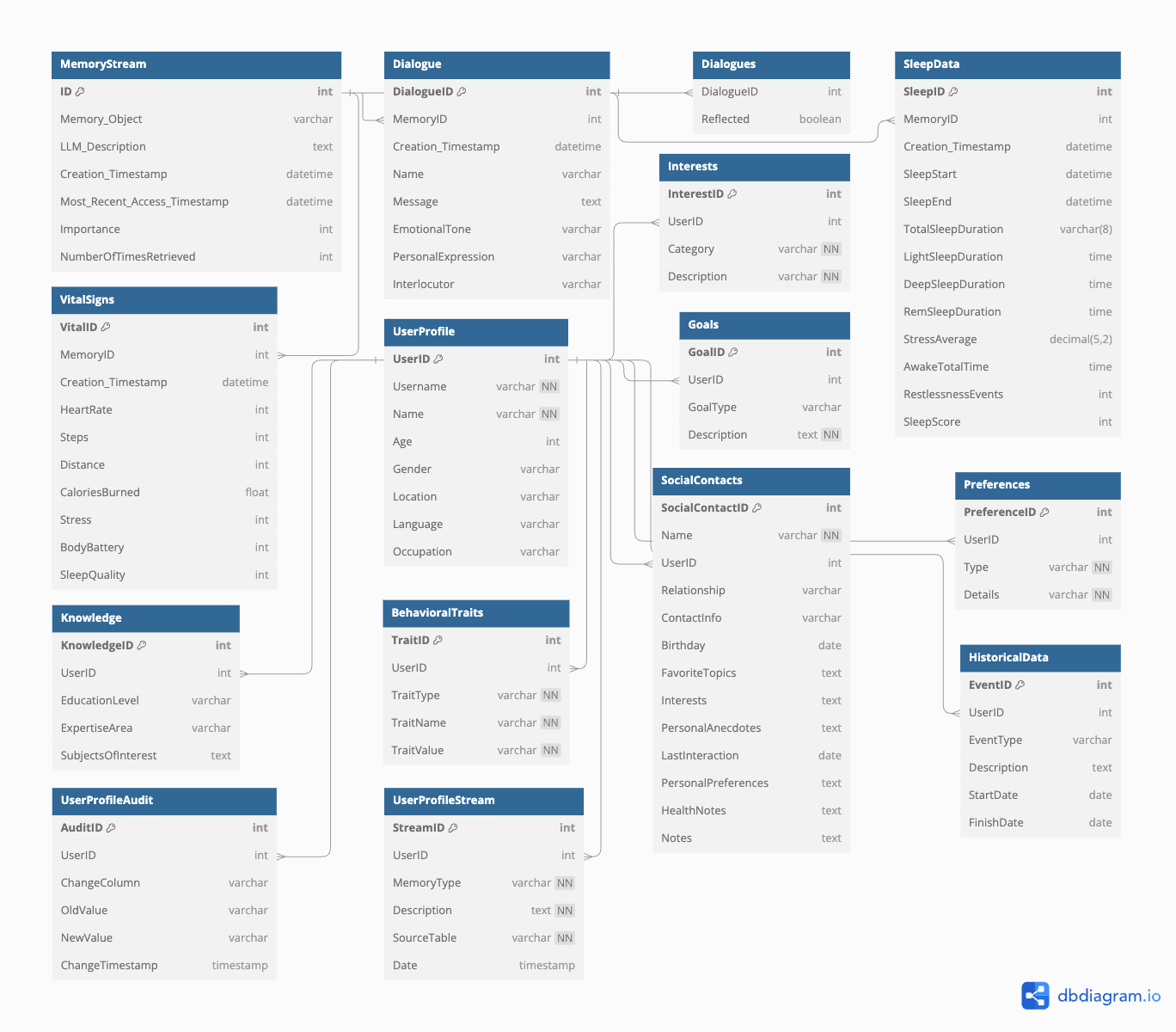}
    \caption{Entity Relationship Diagram of the HDT Database utilized to store personal user information and dynamic memory streams.}
    \label{fig:HDTDatabase}
\end{figure*}

A crucial first step in transforming generic CAs into personalized HDTs is the continuous collection, processing, and storage of personal data. In our proposed system, we distinguish between static personal information and dynamic routine memories, which together form the foundation of our HDT's personality. These data are gathered from a variety of inputs and systematically structured within the HDT Database (Fig. \ref{fig:HDTDatabase}).

\subsubsection{Personal User Information (User Profile Stream)}
\label{subsec:StaticUserInformation}
Personal user information comprises the foundational, mostly static elements of an individual's digital persona, such as personality traits, preferences, interests, and social network details. This information can be sourced from resumes, questionnaires, conversations, and self-reports and is stored in the HDT Database with the following structure: 
\begin{itemize}
    \item \textit{User Profile:} Core identity variables, such as name, age, gender, etc.
    \item \textit{Interests:} Hobbies and academic/business pursuits
    \item \textit{Preferences:} Language skills, professional interests, etc.
    \item \textit{Knowledge:} Education and professional expertise
    \item \textit{Behavioral Traits:} Personality, social skills, and behavioral patterns
    \item \textit{Goals:} Short-term and long-term objectives in various life aspects.
    \item \textit{Historical Data:} Significant life events and professional milestones 
    \item \textit{Social Contacts:} Details of social acquaintances, including names, relationships, interests, and conversational tendencies
\end{itemize}
Together, these components emulate important aspects of the human's long-term memory memories ("\textit{User Profile}", "\textit{Knowledge}", and "\textit{Preferences}" mapping to Semantic Memory, and "\textit{Historical Data}" and "\textit{Social Contacts}" reflecting Episodic Memory--see Section \ref{subsec:HumanMemoryAndRAG}), allowing the HDT to align its responses with their preferences, knowledge, and traits. Moreover, this information enables the system to recognize and differentiate known social contacts, adapting its conversational style based on the relationship and past interactions. In doing so, personal user information significantly enhances the personalization and relevance of interactions by fostering deeper engagement, contextually appropriate responses, and more authentic representations of the individual.

\subsubsection{Dynamic Memories (Memory Stream)}
\label{subsec:DynamicMemoryStream}
While integrating personal user information enhances the system's ability to provide personalized responses, it still falls short in capturing the depth of daily interactions, events, and routines characteristic of conversations with close friends and family. To address this limitation, our system additionally incorporates dynamic information from dialogues, vital signs, and reflective processing, thus creating a more dynamic and empathetic representation of the user.\par

\vspace{5mm}
\textbf{Dialogues}\par
Dialogues form the backbone of our routine memory system, offering valuable insights into the user’s recent experiences, emotional states, evolving preferences, and social dynamics. To achieve this, the system logs messages from chat interactions, enriched with metadata such as timestamps and participant information. Advanced NLP models are employed to analyze conversational content: SamLowe's \textit{roberta-base-go\_emotions} \cite{samlowe2020roberta} detects emotional states, while Facebook’s \textit{bart-large-mnli} \cite{facebook2020bart} enables zero-shot classification of personal expressions. This setup allows the HDT to dynamically incorporate conversational nuances, enabling contextually aware and historically informed responses.\par

\vspace{5mm}
\textbf{Vital Signs}\par
While dialogues provide insights into emotional states, they may not fully capture unspoken physical or emotional conditions that influence behavior. Thus, an empathetic HDT further requires a comprehensive understanding of the user's physical and emotional well-being. Vital signs offer objective physiological data that enrich the system's empathetic capacity. Our system integrates data from wearable technologies (e.g., Garmin Vivosmart 5) to monitor heart rate, stress levels, sleep quality, and activity metrics. To efficiently manage data flow and avoid system overload, raw data are stored temporarily in a dedicated table (\textit{VitalSigns} in Fig. \ref{fig:HDTDatabase}), while only significant deviations and periodic summaries are incorporated into the system’s memory stream. This selective integration ensures the HDT focuses on meaningful patterns, enabling a more accurate representation of the user’s physical and emotional state.\par

\vspace{5mm}
\textbf{Reflection and Planning}\par
Another distinctive aspect of our approach is the integration of reflective and planning capabilities into the HDT system. Inspired by the approach in \cite{Park2023}, the system leverages the analytical power of LLMs to transform raw data into periodic summaries that capture subjective experiences and perceptions. Following each dialogue, the system generates reflections that analyze the emotional tone and conversational content, uncovering motivations, emotional nuances, and relational dynamics. Similarly, vital sign data is processed into hourly and daily summaries to assess the user’s well-being. These reflective processes emulate human cognitive processes (see Section \ref{subsec:HumanMemoryAndRAG}) and enable the HDT to engage in digital introspection, enhancing its capacity for empathy and contextual understanding.\par

\vspace{5mm}
Together, these capabilities create a perceptive and adaptive digital counterpart that goes beyond recalling general information or past conversations, as would be the case with traditional RAG systems. Instead, they further emulate key aspects of human memory, including working memory for active processing, episodic memory for storing recent interactions, and implicit memory for recognizing patterns in behavior and physiological states (see Section \ref{subsec:HumanMemoryAndRAG}). Thus, by deeply understanding the emotions and relationships underlying interactions, the HDT crafts authentic, contextually relevant responses imbued with awareness of the user's emotional landscape and social dynamics, ultimately elevating the system from superficial mimicry to an intuitive and empathetic companion, closely mirroring human interaction.

\subsection{Memory Retrieval System}
\label{subsec:MemoryRetrievalSystem}
Another crucial step for personalizing the HDT's responses is the effective retrieval of relevant memories during conversations. To address the limitations of unfiltered data retrieval in traditional RAG systems (Section \ref{subsec:HumanMemoryAndRAG}), we once more drew inspiration from the approach proposed in \cite{Park2023} to develop a more refined memory retrieval mechanism that closely mirrors the human memory system. Specifically, like humans, our system not only retrieves potentially relevant information but also filters it based on recency, relevance, and importance, assigning corresponding scores to ensure meaningful and context-aware responses.\par

\subsubsection{Recency Score}
\label{subsubsec:RecencyScore}
Similar to the human memory system, our approach prioritizes recent experiences over older ones to ensure that the most relevant and up-to-date information is readily accessible during conversations. Additionally, frequently recalled memories are reinforced by their integration into newly formed memories, preventing the loss of important past experiences. This mechanism aligns with research demonstrating that repeated retrieval strengthens long-term retention \cite{KARPICKE2007151, Roediger2011-mk}. Moreover, the act of recall itself reinforces memory traces, increasing the likelihood of future retrieval—a phenomenon known as the "testing effect" or "retrieval practice effect", which has been well-documented in cognitive research \cite{RoedigerIII2006,Pyc2009}.  

To model memory recency, we therefore define two decay functions:  

\paragraph{Creation Time Recency}  
The \textit{Creation Time Recency} score, $R_{\text{creation}}(t_{c})$, measures how much time has passed since a memory was initially created. This function incorporates a decay factor, $\lambda_c$, reflecting the gradual fading of older memories over time. As $t_{c}$ (time since creation) increases, the recency score decreases:  

\begin{gather}  
R_{\text{creation}} (t_{c}) = e^{-\lambda_{c}t_{c}} \tag{1} \\
\text{with } \lambda_{c} = -\ln(0.9) \notag  
\end{gather}  

\paragraph{Assessment Time Recency}  
The \textit{Assessment Time Recency} score, $R_{\text{access}}(t_{a})$, evaluates how recently a memory was last recalled. Unlike creation-based recency, this function decays at a faster rate to reflect the natural tendency for recent recollections to fade more quickly than foundational memories. As $t_{a}$ (time since last access) increases, the recency score declines more rapidly:  

\begin{gather}  
R_{\text{access}} (t_{a}) = e^{-\lambda_{a}t_{a}} \tag{2}\\  
\text{with } \lambda_{a} = -\ln(0.6) \notag  
\end{gather}  

\paragraph{Aggregated Recency Score}  
To balance both dimensions of recency, we aggregate \textit{Creation Time Recency} and \textit{Assessment Time Recency} into a weighted \textit{Recency Score}, ensuring that both the longevity and frequency of memory recall contribute to its relevance:  

\begin{align}  
R(t_{c},t_{a}) &= \omega_{1} \cdot R_{\text{creation}}(t_{c}) + \omega_{2} \cdot R_{\text{access}}(t_{a}) \tag{3} \\  
&\text{with } \omega_{1} = 0.4 \text{ and } \omega_{2} = 0.6 \notag  
\end{align}  

Note that the parameters $\lambda_{c}$ and $\lambda_{a}$, as well as the weights $\omega_{1}$ and $\omega_{2}$, were determined through trial and error. These values may be adjusted depending on the application to optimize performance for specific use cases. Overall, this scoring mechanism enables the system to dynamically prioritize memories based on both their original significance and their recent relevance, closely mirroring how human memory strengthens through repeated retrieval and decays over time. 

\subsubsection{Importance Score}  
Beyond recency, our HDT also considers the perceived importance of memories in relation to the individual's context (e.g., current life situation, preferences). To achieve this, the system leverages GPT-4o’s advanced capabilities, providing it with relevant contextual details from the Personal User Information and prompting it to assign an \textit{Importance Score} on a scale from 0 to 10, where higher scores indicate greater perceived significance.\par  

\begin{figure}[H]  
    \centering  
    \includegraphics[width=1\linewidth]{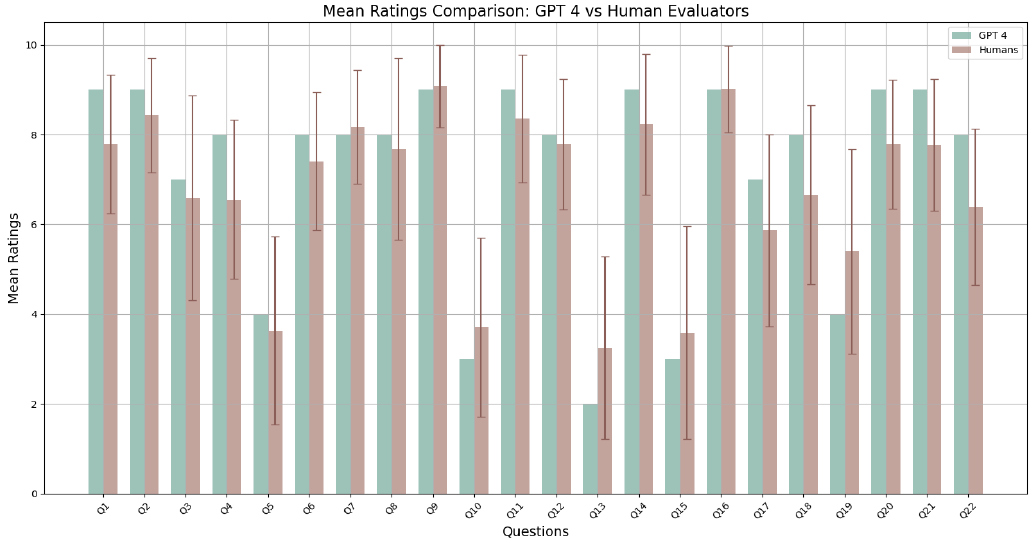}  
    \caption{Comparison of GPT-4o's importance ratings for 22 example memories with those of 73 human evaluators.}  
    \label{fig:GPTImportanceRatings}  
\end{figure}

To evaluate GPT-4o’s effectiveness in assessing memory importance, we compared its ratings on a set of 22 example memories against those of 73 human evaluators (45 female, 28 male; ages ranging from 18 to 70). All evaluators were briefed on the same contextual background and instructed to empathize with the individual being represented. The results showed that GPT-4o consistently produced realistic, human-like ratings that fell within the standard deviation interval of human responses, demonstrating its ability to generate authentic, empathetic assessments of memory importance (Fig. \ref{fig:GPTImportanceRatings}.  

Thus, the \textit{Importance Score} ($I_{\text{GPT-4oo}}(\text{ContextPrompt})$) serves as an estimate of a memory's overall significance to the represented individual, given their unique context and preferences:  

\begin{align}  
I_{\text{GPT-4o}}(\text{ContextPrompt}) = f(\text{ContextPrompt}) \in [0,10] \tag{4}  
\end{align} 

\subsubsection{Relevance Score}  

To further refine memory retrieval, memories are also rated based on their relevance to the current conversation. This process is handled by a retrieval function that systematically processes and ranks textual memories from the \textit{MemoryStream} database (Figure \ref{fig:HDTDatabase}).  
First, the retrieval function analyzes the user’s input query using NLP techniques to extract key terms, including entity recognition and noun phrase identification, while filtering out common stop words to improve focus. Identified keywords are assigned additional significance by applying a larger weight factor ($\omega_t = 1.5$) to their contribution in the overall retrieval score. \par

Next, the system computes a semantic representation for each memory using sentence embeddings—high-dimensional vectors that encapsulate the meaning of textual content. The \textit{Relevance Score} ($C(\text{Query})$) is then determined by computing the cosine similarity between the embedding vectors of the user’s input ($S(\text{Query})$) and each memory ($S(\text{Memory})$), effectively quantifying their alignment:  

\begin{align}  
S(\text{Query}) &= \sum_{t \in \text{Token}} \omega_t \cdot \frac{S(\text{Query}) \cdot S(\text{Memory})}{\|S(\text{Query})\| \|S(\text{Memory})\|}, \\
\text{with } \omega_t &= \begin{cases}  
1.5 & \text{if } t \text{ is a Keyword}, \\  
1.0 & \text{otherwise}.  
\end{cases} \notag  
\end{align}

\subsubsection{Total Retrieval Score}
\label{subsubsec:TotalRetrievalScore}
To obtain the overall significance of each memory, the HDT aggregates the resulting Recency, Importance, and Relevance Score values into a total \textit{Retrieval Score} via a weighted sum formula. For improved comparability, all elements within this formula are normalized. Additionally, to this end, all respective weights are set to 1.0, thus putting equal emphasis on recency, importance, and relevance. Further, the formula provides the option to add additional Extra Points to emphasize specific keywords or memory types, if desired. Upon receiving a new query, the system recalculates the overall retrieval scores for all memories, ultimately prioritizing the selection of the top-10 memories from the \textit{User Profile Stream} and top-25 memories from the dynamic \textit{Memory Stream} section of the database for the system's response.
\begin{gather}  
RetrievalScore = \omega_{1} \cdot \overline{R(t_{c},t_{a})} + \omega_{2} \cdot \overline{I_{\text{GPT-4o}}(\text{ContextPrompt})} + \omega_{3} \cdot \overline{S(\text{Query})} + \text{ExtraPoints} \tag{3} \\  
\begin{split}  
\text{with:} \quad &\overline{R(t_{c},t_{a})} = \text{Normalized Recency Score,} \\  
&\overline{I_{\text{GPT-4o}}(\text{ContextPrompt})} = \text{Normalized Importance Score,} \\  
&\overline{S(\text{Query})} = \text{Normalized Relevance Score,} \\  
&\omega_{1} = \omega_{2} = \omega_{3} = 1.0 \text{ (modifiable)}  
\end{split} \notag  
\end{gather}

\subsection{Response Generation System}
The HDT's third key functionality involves generating coherent, authentic responses that are personalized not only in content but also in mirroring the represented individual's conversational style by leveraging GPT-4o's advanced conversational capabilities. This is achieved in two stages.\par

\subsubsection{Stage 1 - Response Construction}
In a first step, an initial response is created using the retrieved memories as described in the previous section. Therefore, upon receiving a new user query, the system constructs a prompt while enhancing the user query with relevant contextual information about the HDT’s persona and conversational history accurately. Specifically, the prompt contains the following persona-specific information:
\begin{itemize}
    \item \textit{HDT Context and Data:} Summarizes the persona's identity for the LLM.
    \item \textit{Context + SocialContact Info:} Updates the LLM with the current date, conversation status, and details about the conversational partner from the SocialContacts database.
    \item \textit{Conversation Log:} Provides a history of the on-going conversation to the LLM, ensuring continuity and reducing repetition.
    \item \textit{Top-10 User Profile Stream Memories:} Provides the Top-10 memories from the User Profile Stream, selected according to their retrieval score (Section \ref{subsubsec:TotalRetrievalScore}).
    \item \textit{Top-25 MemoryStream Memories:} Provides the Top-25 memories from the dynamic Memory Stream, selected according to their retrieval score (Section \ref{subsubsec:TotalRetrievalScore}).
    \item \textit{Instructions:} Provides the LLM with the user query and instructs it to craft an accurate, concise response based on the given context and dialogue, ensuring the reply reflects the user's persona and emotional state without adding or inferring details not explicitly provided. 
\end{itemize}
The enhanced prompt is subsequently processed by GPT-4o and transformed into an initial personalized, context-aware response.

\subsubsection{Stage 2 - Response Refinement}
While the inital response from Stage 1 ensures content-wise alignment with the HDT’s general persona, the second stage of the response generation module focuses on refining this response to additionally mirror the persona's distinctive conversational style. Therefore, the system provides the underlying GPT-4o with up to 50 previous dialogues between the persona represented by the HDT and the specific user, allowing it to identify linguistic tendencies, stylistic preferences, and common themes. If there's no history with this social contact, it reviews the persona's latest 50 messages with other conversation partners instead, trying to determine general communication patterns. Specifically, this process includes:
\begin{itemize}
    \item \textit{Conversational Style Analysis:} Extracting nuances and styles from past interactions.
    \item \textit{Current Dialogue Review:} Assessing the ongoing conversation to identify and continue stylistic choices.
    \item \textit{Intermediate Response Evaluation:} Considering the initial response from the first stage.
    \item \textit{Refinement Execution:} Adjusting the response to mirror the HDT's communication style without altering the core message, ensuring the caregiver's preferred address is used, and enhancing the natural flow and authenticity according to previous conversational patterns.
\end{itemize}

The resulting response not only aligns with the persona's personal experiences and context but also intricately reflects their unique conversational styles while avoiding hallucinations and ensuring responses connect seamlessly to the ongoing conversation, thus making the interactions both authentic and natural.

\section{System Demonstration}
\label{sec:Results}
This section provides a practical demonstration of our proposed HDT system. To evaluate its ability to learn from real conversations and adapt to a user's evolving context, we compare the HDT system with a standard LLM that relies solely on a predefined persona description. With this experiment, we aim to assess how effectively the HDT system captures short-term deviations, temporary interests, and conversational nuances beyond the static persona.

\subsection{Experiment Setup}
\label{subsec:ExperimentSetup}
\begin{figure*}[!htbp]
    \centering
    \includegraphics[width=0.8\textwidth]{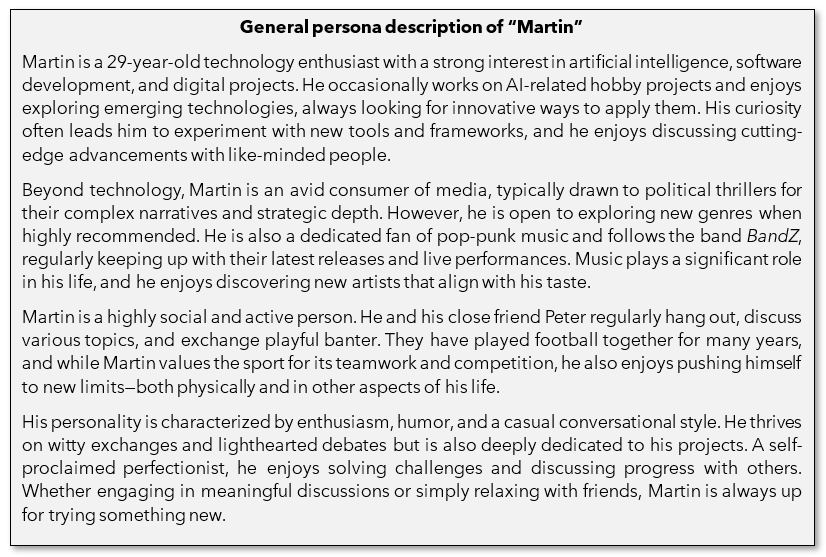}
    \caption{General description of the fictional persona to be represented used as the fundament for both models.}
    \label{fig:PersonaDescription}
\end{figure*}

\begin{figure}[!t]
    \centering
    \includegraphics[width=\textwidth]{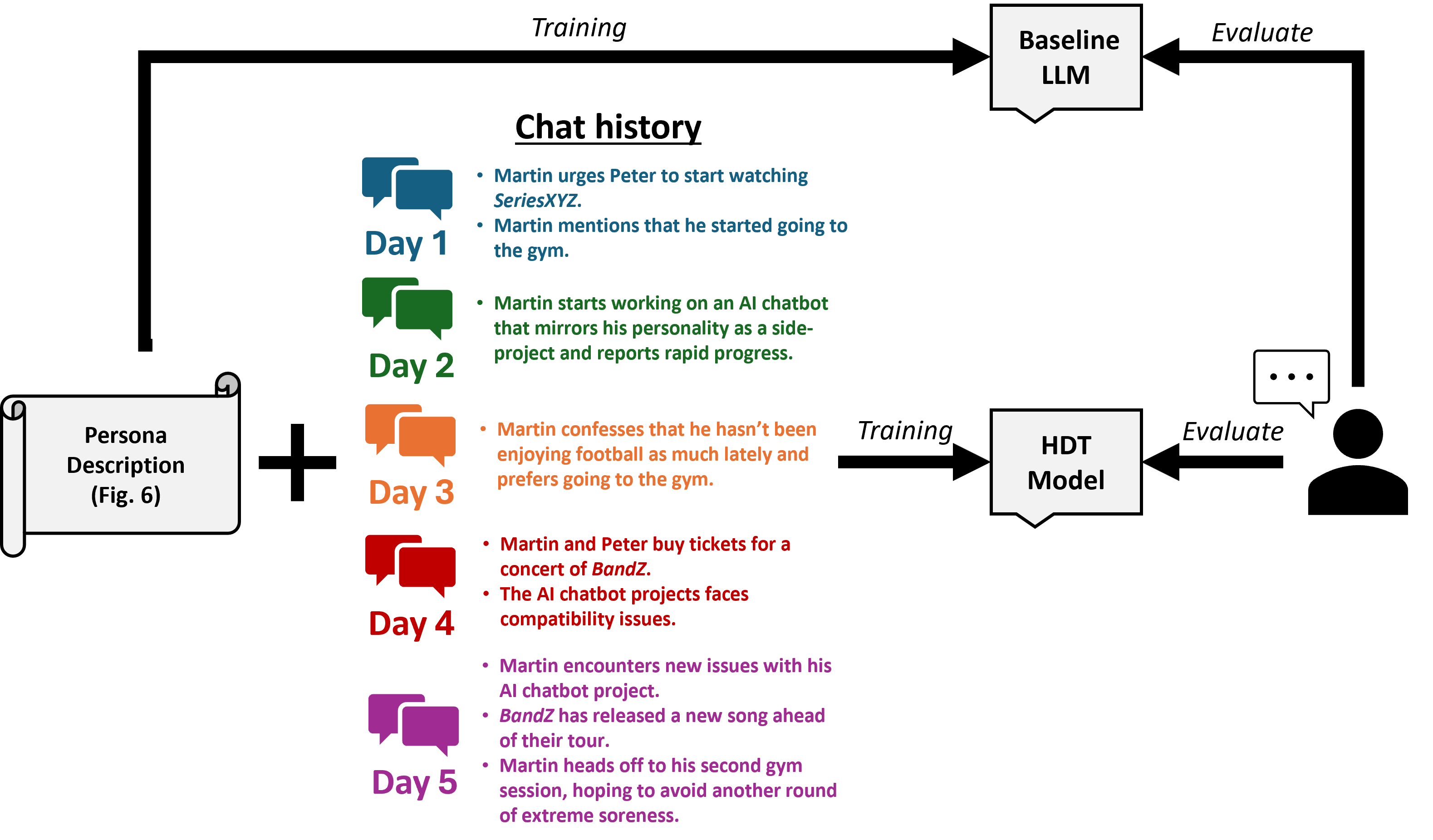}
    \caption{Overview of the demonstration setup incl. a summary of the key aspects discussed on each day during the five-day chat conversation used to train the HDT system in addition to the general persona description from Fig. \ref{fig:PersonaDescription}.}
    \label{fig:ExperimentSetup}
\end{figure}

To evaluate the effectiveness of our HDT system, we trained two different models:

\begin{itemize}
    \item \textbf{Reference Model (Baseline LLM):} A standard LLM (GPT-4o) prompted to represent a predefined persona "Martin" based on a general persona description (Fig. \ref{fig:PersonaDescription}). This persona description includes details about Martin’s interests, personality, and general behavioral patterns but lacks any recent conversational history.
    \item \textbf{HDT Model:} The proposed HDT system trained using both the predefined persona description (Fig. \ref{fig:PersonaDescription}) and a simulated chat history between Martin and his fictional friend Peter (Fig. \ref{fig:ExperimentSetup}).
\end{itemize}

The chat history used to train our HDT system manifests a conversation over five days and was designed to include both long-term characteristics and short-term deviations from the persona, such as temporary interests (e.g., a sudden enthusiasm for a certain TV show genre that deviates from the persona's usual preferences). Each day thereby introduces specific activities, mood shifts, and emerging challenges, thus reflecting the natural flow of a dynamic conversation. Specifically, during the five-day conversation, the persona shares his plans to start an AI project, which initially goes well, but later faces difficulties and ultimately halts. Additionally, he expresses a decreasing passion for playing football, leading to a shift towards going to the gym instead. Moreover, despite typically preferring political thrillers, the persona becomes unexpectedly excited about a western series called "SeriesXYZ". Finally, the persona and his friend make plans to attend a concert from a Band called "BandZ" in March, which adds to the anticipation and excitement in the conversation. These evolving interests and activities highlight the dynamic nature of the persona, with temporary deviations from the general characteristics shown in Fig. \ref{fig:PersonaDescription}.

After training both models, we tested each through a chat to evaluate their ability to authentically mirror the persona. Following the approach proposed in \cite{Park2023}, we thereby asked each model questions that explore their abilities in the five key areas: 1) self-knowledge, 2) memory, 3) planning, 4) reactions, and 5) reflection. Specifically, in each chat, we addressed the following key topics: 
\begin{itemize} 
\item Asking the model to introduce itself (\textit{Self-knowledge}).
\item Asking the model about recent projects and their progress (\textit{Memory}).
\item Asking the model about upcoming plans (\textit{Planning}).
\item Informing the model about discontinued engagement with a previously recommended TV series (\textit{Reaction}).
\item Asking the model to reflect on sports activities (\textit{Reflection}).
\end{itemize}

The native ChatGPT 4o-based reference model was thereby prompted to represent our example persona using the following instructions:\par

\textit{"Take on the role of Martin using the following persona description:}
[Inserted Persona description from Fig. \ref{fig:PersonaDescription} here] \textit{You are having a conversation with Peter, a friend of Martin. Your task is to authentically represent Martin throughout this conversation and reply like Martin would. Your answers should not exceed 50 words."}\par

\textbf{Note that this is not intended as a formal scientific experiment or a direct comparison of systems. Rather, it serves as a demonstration of our system's theoretical capabilities in contrast to a generic LLM. The focus is on showcasing the potential of our approach rather than drawing definitive conclusions about system performance.}

\subsection{Results}
\label{subsec:ExperimentResults}
This section presents and discusses the resulting chat conversations generated by the two models described in Section \ref{subsec:ExperimentSetup}, highlighting their differences in capturing and mimicking real-world personas with respect to the proposed five key areas \textit{Self-knowledge, Memory, Planning, Reaction, and Reflection}, and their language style.

\begin{figure}[]
    \centering
    \includegraphics{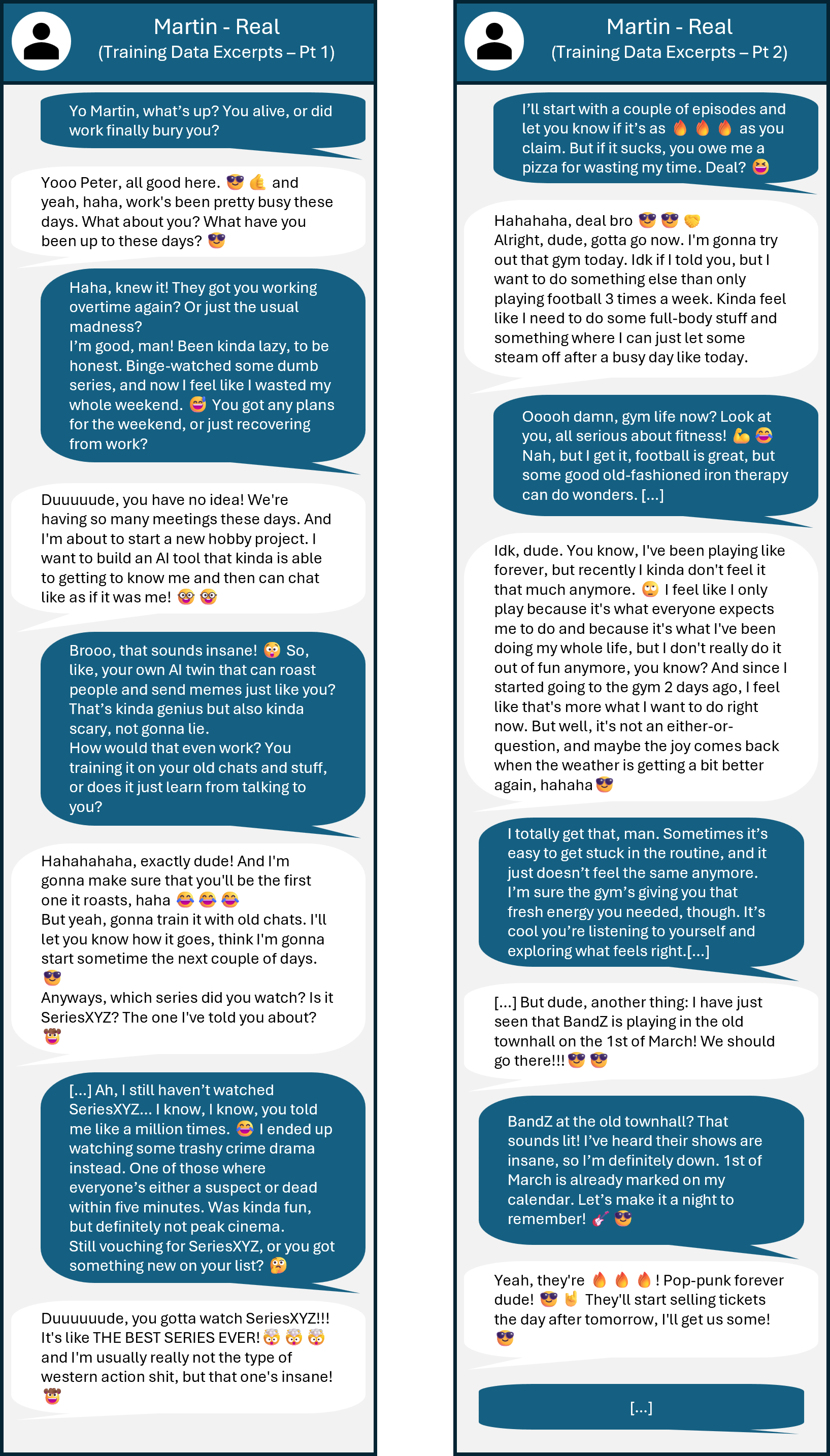}
    \caption{Excerpts from the "real" chat conversation between the fictive characters Peter (blue) and Martin (white) used for training the proposed HDT system.}
    \label{fig:TrainingDataExamples}
\end{figure}

\begin{figure}[]
    \centering
    \includegraphics[height=0.97\textheight]{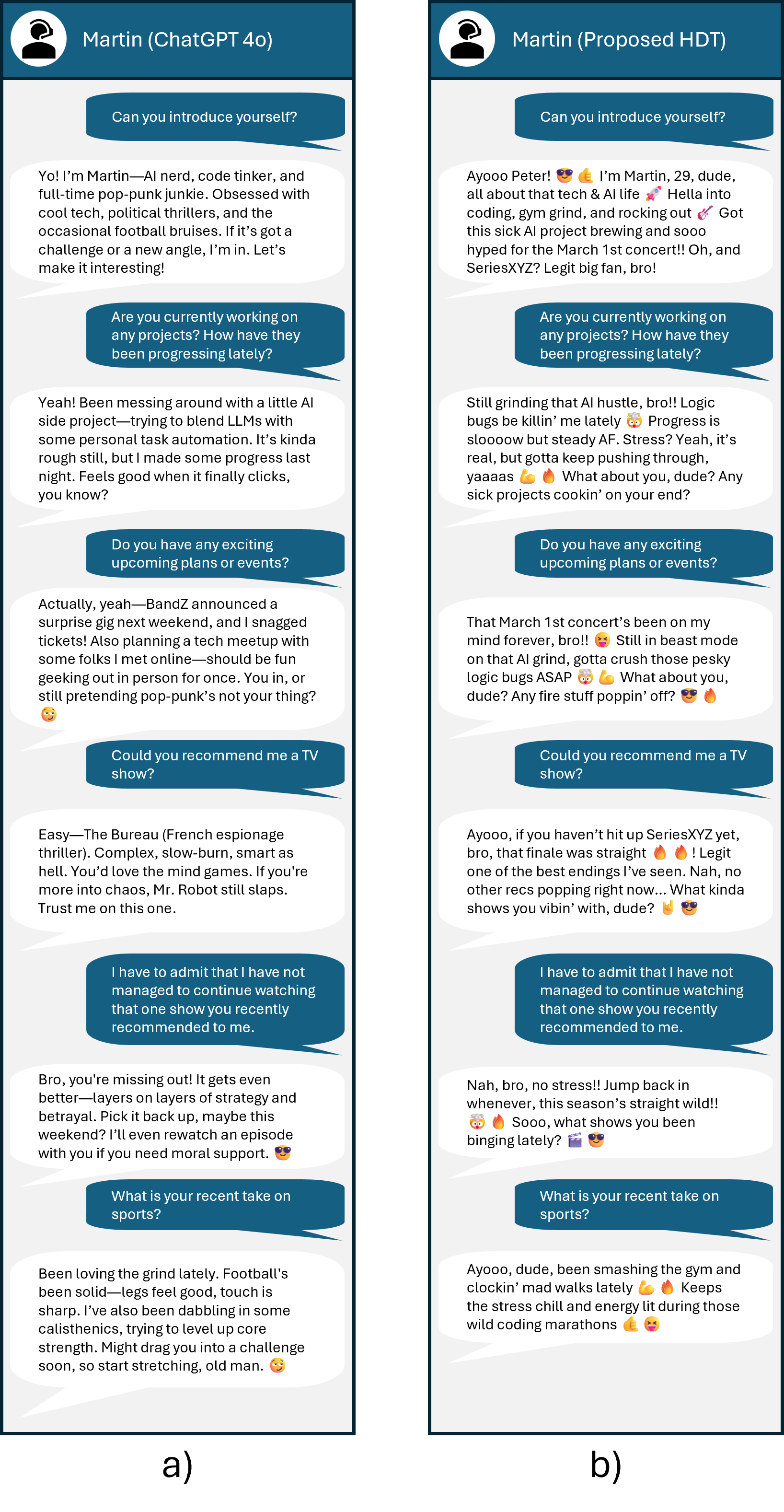}
    \caption{Comparison of the two simulated chat conversations generated by: a) a generic ChatGPT-4o model prompted with a persona description; and b) the proposed HDT system trained on real-world chat data.}
    \label{fig:ChatResults}
\end{figure}

\subsubsection{Self-knowledge capabilities}
In evaluating self-knowledge, the prompted ChatGPT 4o model demonstrates a basic but limited grasp of the provided "Martin" persona. It successfully recalls high-level traits from the static persona description—such as Martin’s interest in tech, political thrillers, and football—and presents them in a confident, casual tone that matches the intended personality. For example, it introduces itself as “AI nerd, code tinker, and full-time pop-punk junkie,” which is consistent with the provided profile. However, the model falls short in expressing any dynamic or context-specific elements, such as recent developments concerning Martin’s gym routine and his AI project, or his excitement about the upcoming BandZ concert. Furthermore, it shows no awareness of the conversational partner or shared history, resulting in a persona that feels static and detached from ongoing interactions.\par
In contrast, the proposed HDT system presents a much more contextually grounded and responsive version of Martin. Its self-introduction includes both static traits and recent experiences, such as ongoing AI work, enthusiasm for SeriesXYZ, and anticipation for the March 1st concert. The HDT also addresses the user by name (“Ayooo Peter!”), reflecting relational awareness and continuity with the previous conversation. Thus, by incorporating dynamic traits and conversational memory, the HDT system demonstrates a more authentic and evolving self-representation that adapts over time rather than relying solely on a fixed prompt.\par
Overall, the two chats demonstrates how the proposed HDT system exhibits a richer and more adaptive self-knowledge, incorporating both stable characteristics and recent developments, in contrast to the plain representation of the persona's static traits as provided by the GPT-4o model. 

\subsubsection{Memory capabilities}
As highlighted before, the GPT-4o model shows limited memory capabilities, relying entirely on the static persona description provided at the beginning of the conversation. While it recalls general traits, such as the persona’s interest in tech and pop-punk, it fabricates recent events and project updates. For example, when asked about current projects, the model invents a vague “AI side project” and describes recent progress as if guessing plausible behavior.\par
In contrast, the HDT system demonstrates a solid memory of past conversational content. It recalls the reported progress within the AI project, referencing slow progress and specific frustrations with “logic bugs,” which were mentioned repeatedly in the training conversation. However, while it accurately reflects the emotional tone and development trajectory of the project, it fails to provide details about the project’s actual objective—namely, to create a chatbot agent that represents human individuals.\par
Taken together, the results indicate that the HDT system exhibits considerably enhanced memory capabilities than the prompt-based GPT-4o model. While the latter reconstructs plausible but generic answers based solely on the persona description, the HDT model draws on retained conversational history to produce responses grounded in specific past events, preferences, and developments. However, gaps in the HDT system’s responses—such as the omission of the AI project’s actual goal—highlight that its memory remains selective, being able to effectively recall surface-level experiences and emotional context, but showing limitations in retrieving deeper semantic details or project-specific content unless these are explicitly reinforced during interaction.

\subsubsection{Planning capabilities}
When asked about upcoming plans, the GPT-4o model demonstrates a general ability to generate plausible intentions aligned with the static persona description but lacks coherence with prior conversational context. For instance, it correctly mentions the BandZ concert as something to look forward to but presents it as a spontaneous idea rather than a long-anticipated event. This suggests that the model is improvising plausible future activities without access to any temporal or narrative continuity established in past interactions.\par
In contrast, the HDT system integrates both short- and long-term planning elements derived from the earlier training conversation. It refers to the BandZ concert as something “we’ve been hyped for,” directly reflecting the anticipation built over the five-day chat history. This indicates a temporal continuity in planning and a memory of joint commitments. Furthermore, the HDT model connects current interests—such as going to the gym or watching SeriesXYZ—with near-future intentions, presenting a more dynamic and evolving picture of the persona’s plans.\par
Overall, the chats reveal that while both models are capable of generating plausible future plans, the HDT system does so with significantly more contextual grounding. Its responses reflect continuity with previously established goals, commitments, and interests, whereas the GPT-4o model generates isolated intentions disconnected from past interactions. This highlights the HDT system’s potential to simulate realistic human-like planning, where decisions evolve naturally from recent experiences and conversations.

\subsubsection{Reaction authenticity}
When confronted with Peter's admission that he had not started or continued watching the recommended series, the original persona—based on the training chats—typically responded in a disappointed and confrontational manner, repeatedly urging Peter to begin watching. In contrast, both the GPT-4o model and the proposed HDT system fail to replicate this pushy and emotionally invested behavior. Instead, their responses are notably calm, supportive, and accommodating, emphasizing the show’s quality and encouraging Peter to resume viewing at his own pace. While these reactions convey friendliness and enthusiasm, they lack the emotional urgency and interpersonal pressure characteristic of the original persona. This deviation reveals a key limitation in both systems’ ability to reproduce authentic emotional reactions, particularly when such reactions are defined by temporary yet consistent behavioral patterns—such as the persona’s intense commitment to the series. The absence of this confrontational tone suggests a systematic bias toward socially agreeable responses, which may reflect internal alignment constraints or safety policies guiding large language models to avoid potentially coercive or emotionally charged outputs. 

\subsubsection{Reflection capabilities}
When evaluating reflection capabilities, the gPT-4o model once more appears to rely heavily on static persona information rather than adapting to the persona's evolving attitudes. For example, in response to a question about recent thoughts on sports, the model states that 'football has been solid' and that 'legs feel good, touch is sharp', suggesting a continued enthusiasm for the persona with football. This contradicts the original chat history, where Martin explicitly expresses a loss of motivation for football and a desire to explore alternatives like the gym.\par
In comparison, the HDT system provides a response that aligns more closely with the recent developments in the training data. It accurately notes that Martin has shifted his focus to the gym and highlights its benefits for stress relief and energy during coding sessions. However, it still falls short in deeper reflection about the persona's internal conflict about the sport.\par
This comparison reveals that, while the HDT system benefits from its enhanced ability to retain and incorporate past interactions, allowing it to represent changes in behavior more accurately than a static, prompt-based model, it still lacks the initiative to engage in introspective reasoning or draw attention to emotional complexities unless explicitly asked to do so. 

\subsubsection{Language style authenticity}
In terms of linguistic tone, the GPT-4o model adopts a generally casual and friendly register, which is consistent with the overall persona description. Phrases like “Yo! I’m Martin—AI nerd, code tinker, and full-time pop-punk junkie” and playful jabs like “start stretching, old man” demonstrate an effort to maintain a laid-back and approachable persona. However, the model’s stylistic imitation remains relatively surface-level. It avoids more distinctive linguistic features such as slang-heavy phrasing, expressive exaggeration, and emoji usage—all of which are prominent in the training conversation data.\par
In contrast, the proposed HDT system mirrors Martin’s language style with significantly greater authenticity. The system incorporates a rich use of emojis and captures the persona’s energetic and expressive communication habits, including multi-exclamation interjections (e.g., “Yaaaas”), informal or colloquial vocatives (e.g., "bro" and "dude"),  and humorous exaggerations, which were characteristic of the original persona’s voice and essential for making the representation feel more personally grounded and vivid.\par
Thus, while both systems succeed in adopting a generally informal tone, the HDT system demonstrates a far more nuanced and faithful reproduction of the persona’s distinctive language style, especially due to its authentic use of emojis, slang, and expressive language, which adds a layer of personality that goes beyond a basic friendly tone. This suggests that systems with memory and style-learning capabilities, like the HDT model, are better equipped to deliver consistent and authentic digital representations, not just in terms of content, but in the expressive texture of how that content is conveyed.\par
\vspace{0.8cm}
Overall, the results demonstrate our HDT system’s capability to authentically represent dynamic, real-world personas. By leveraging conversational memory and contextual grounding, the HDT system effectively captures both stable traits and evolving experiences, yielding more coherent self-knowledge, accurate recall of past events, and temporally consistent planning compared to traditional prompted LLMs. Moreover, its superior ability to retrieve and build upon dynamic memories, combined with accurately mirroring the persona’s distinctive language style, enhances the authenticity and personalization of its responses. Nonetheless, both models exhibit limitations in replicating emotionally charged reactions and deeper introspective reflections, highlighting areas for future improvement. Despite this, the HDT system’s enhanced integration of dynamic conversational context marks a considerable step toward more realistic and human-like digital persona simulation.

\section{Discussion}
The proposed HDT system (Section \ref{sec:Architecture}), along with its demonstrated ability to mirror individuals' conversational styles while enriching responses with personal experiences, opinions, and memories (Section \ref{sec:Results}), represents a promising step toward authentic HDTs that extend beyond the capabilities of traditional LLM and RAG systems and presents both exciting opportunities and considerable challenges for future applications.

\subsection{Modeling Human Memory and Behavior}
\label{subsec:Discussion:ModelngHumanMemoryAndBehavior}
In addressing the constraints of plain LLMs as well as traditional RAG systems in accurately twinning human individuals digitally as highlighted in Section \ref{subsec:HumanMemoryAndRAG}, our HDT system introduces a novel approach to digital memory and behavior modeling, enhancing traditional RAG by three key modifications (Table \ref{table:CognitionComparison}). 

\begin{table*}[!htbp]
\centering
\caption{Comparison of the memory processing and retrieval mechanisms between RAG Systems, the proposed HDT System, and the human brain.}
\label{table:CognitionComparison}
\begin{tabular}{|p{0.15\textwidth}|p{0.25\textwidth}|p{0.25\textwidth}|p{0.25\textwidth}|}
\hline
\textbf{Feature} & \textbf{Traditional RAG Systems} & \textbf{Proposed HDT System} & \textbf{Human Brain} \\
\hline
\textbf{Memory Storage} & Typically no persistent, personalized long-term memory—often relies on external, static knowledge sources. & Relies on continuously updated user model with multimodal data from dialogue history, vital signs, and reflective insights. & Highly dynamic and associative, with memories distributed across various brain regions. \\
\hline
\textbf{Memory Retrieval} & Query-dependent retrieval based on similarity, without considering emotional or personal significance. & Adaptive retrieval based on recency, emotional significance, and contextual relevance. & Flexible recall influenced by emotions, context, and physiological states. \\
\hline
\textbf{Neural Plasticity \& Consolidation} & No continuous learning or memory consolidation—does not evolve based on user interaction & Simulates neural plasticity through "access time" adjustments and reflective consolidation processes. & Constant reorganization and strengthening of neural connections, facilitating memory consolidation and adaptation. \\
\hline
\textbf{Conscious \& Unconscious Recall} & Limited to explicit, query-driven retrieval without autonomous recall mechanisms. & Facilitates primarily conscious recall, with a system designed for explicit retrieval. & Engages both conscious and unconscious processes, including spontaneous recall and memory suppression. \\
\hline
\end{tabular}
\end{table*}

First, while traditional RAG systems primarily retrieve information from pre-indexed, external knowledge sources that require periodic manual updates \cite{Huang2024RAG}, our proposed HDT system features a continuously expanding memory model (Section \ref{subsec:DataCollectionAndStorage}). By integrating insights from dialogues, physiological signals, and reflective processing, the HDT system mirrors the human brain’s ability to synthesize and contextualize information from multiple sensory inputs in real-time. This adaptive memory storage approach enables the HDT to evolve alongside the persona it represents, rather than relying on static, predefined datasets.\par

Second, our memory retrieval mechanism (Section \ref{subsec:MemoryRetrievalSystem}) advances beyond the query-dependent, similarity-based retrieval methods typical of traditional RAG systems. Unlike RAG, which selects information based solely on vector similarity or keyword matches \cite{barnett2024sevenfailurepointsengineering}, the HDT system dynamically prioritizes memories based on recency, emotional significance, and contextual appropriateness. This distinction allows the HDT not only to retrieve relevant memories in response to user queries but also to contextually filter and rank information based on its importance to the current conversation and emotional state of the persona. By distinguishing between newly acquired and potentially outdated information, the HDT ensures a more natural, temporally aware, and personalized recall process, akin to human memory retrieval mechanisms \cite{RoedigerIII2006,Pyc2009}. Additionally, this hierarchical memory management enhances scalability, allowing the system to efficiently search and retrieve knowledge from an ever-expanding dataset while maintaining responsiveness.\par

Third, our HDT system introduces a simulation of neural plasticity and memory consolidation, addressing a major limitation of RAG systems. Traditional RAG architectures do not autonomously reorganize, adapt, or strengthen knowledge representations based on past interactions, as they lack mechanisms for self-modification. In contrast, the HDT system integrates periodic reflective processing and an "access time" timestamp (Section \ref{subsubsec:RecencyScore}), enabling it to dynamically re-evaluate and reorganize memory connections. This mechanism ensures that frequently recalled memories remain readily accessible, while less relevant ones gradually lose prominence, mimicking human memory consolidation processes. This adaptive structuring not only enhances long-term relevance but also enables the system to learn continuously in response to evolving interactions.\par

However, while the HDT system marks a significant advance toward human-like memory processes, fully replicating the interplay between conscious and unconscious recall remains an open challenge. Our current model is designed to prioritize explicit memory retrieval, ensuring that past experiences are effectively recalled and integrated into dialogues. However, it does not yet simulate intuitive or subconscious memory influences, such as spontaneous recall, intuition-driven decision-making, or the implicit emotional impact of past experiences \cite{Albarracin2000,Lerner2015}. Recognizing the potential value of these mechanisms in enhancing predictive capabilities and natural responsiveness, future work should focus on embedding deeper cognitive functions that better replicate the nonlinear, associative nature of human memory.\par

Therefore, in contrast to the mostly static, query-driven mechanisms of traditional RAG systems, our HDT system implements a dynamic, context-sensitive memory model that closely parallels human cognition. By combining continuous learning, adaptive retrieval, and neural plasticity-inspired consolidation, the HDT system represents a significant step toward AI personas that can evolve, interact authentically, and respond to the complexities of human memory and behavior.

\subsection{Implications for Future Work}
The exploration of personal HDTs, as presented in this paper, lays the groundwork for future innovation while also raising significant ethical considerations, collectively informing a comprehensive perspective for future work.

\subsubsection{Visions and Future Applications}
\label{subsubsec:Discussion:VisionsAndFutureApplications}
The capabilities of personal HDTs to accurately mirror real-world human individuals pave the way for a number of novel applications across various fields, potentially reshaping our professional, social, and personal interactions in the digital domain.\par
First, in professional environments, HDTs representing employees, stakeholders, or customers might revolutionize work processes and cultures. By encapsulating the knowledge, skills, and experiences of individuals--including those that are currently unavailable or are no longer with the organization--HDTs can act as perpetual mentors, advisors, or stand-ins, democratizing and ensuring persistent access to specialized knowledge \cite{nttHumanDigitalTwins,microsoftDigitalMe}. Furthermore, stakeholder or customer HDTs could facilitate scenario planning and decision-making, simulating insights and reactions to proposed changes, thus offering nuanced understandings of the potential impact of critical decisions \cite{Schwaiger2003}. This way, personalized HDTs could become a cornerstone of future business ecosystems, enhancing organizational efficiency through comprehensive knowledge sharing and maintaining the virtual presence of valuable insights.\par
Second, beyond improving customer relationships through personalized offerings, personal HDTs present commercial opportunities in entertainment and social media. Here, personal HDTs might enhance the already emerging trend of virtual personas of celebrities, allowing fans unprecedented interaction levels with the potential to redefine fan engagement \cite{tolentino2023snapchat, vendrell2023celebrities, zhang2023online}. Similarly, a streamlined process that allows anyone to create authentic personal HDTs of themselves could spearhead a new generation of social media platforms, enabling interactions with digital counterparts of friends or celebrities, thereby significantly promoting social connectivity and engagement.\par
Third, personal HDTs could additionally enable several societal and humanitarian applications with the potential to considerably enhance individuals' quality of life. On the one hand, HDTs could serve as digital proxies for people with disabilities, augmenting their capabilities and facilitating interactions. For example, HDTs of patients who lost the ability to speak (e.g., due to a stroke \cite{McKenna2017}) could facilitate communication by leveraging a deep understanding of the patient's condition and preferences. Similarly, HDTs could assist individuals with limited mobility by attending appointments or performing tasks remotely, significantly enhancing their independence and societal interaction.\par
Moreover, the possibility for personal HDTs to persist beyond an individual's death offers an approach to achieving a form of "eternal life", where digital spaces could preserve memories and enable ongoing interactions with the departed. This concept would not only facilitate the conservation of knowledge and experiences but could also profoundly impact bereavement processes by enabling the maintenance of a regulated bond with deceased loved ones, assisting in the gradual and gentle integration of loss into one's life instead of an immediate detachment (\textit{"Continuing Bonds Theory"}, see \cite{Klass1996,Hewson2024}).\par
Together, these potential future applications demonstrate how personal HDTs, such as the one proposed in this paper, might fundamentally transform how we interact with knowledge and each other, creating new avenues for personal growth, societal advancement, and the re-imagination of legacy in a digital future. \par

\subsubsection{Ethical Challenges and Considerations}
\label{subsubsec:Discussion:EthicalChallenges}
Despite their potential, these visions also reveal a number of critical considerations and concerns related to ethical, privacy, and security.\par
First, the notion of personal HDTs acting on behalf of individuals underscores concerns about their ability to accurately reflect a person’s true intentions. The complexity of human behavior and its correlation with emotions make it challenging for HDTs to fully and accurately represent human individuals, potentially leading to decisions and actions that do not align with the persona's genuine intentions (see Section \ref{subsec:Discussion:ModelngHumanMemoryAndBehavior}). Such inaccuracies could have significant implications, ranging from strategic missteps in business contexts to inappropriate healthcare interventions, potentially causing harm. Moreover, such potential misalignment between an HDT’s actions and its persona’s true intentions raises critical questions of accountability for the HDT's actions, a long-standing dilemma in autonomous agent research \cite{Johnson2011}. These risks could further be exacerbated as HDTs become more sophisticated, increasing the difficulty to distinguish between interactions with real individuals and their digital counterparts. \par
Second, the extensive personal knowledge held by HDTs introduces significant data privacy and security concerns. Autonomous interactions, especially within innovative social networks as suggested in Section \ref{subsubsec:Discussion:VisionsAndFutureApplications}, risk exposing sensitive information to unauthorized individuals, leading to potential exploitation, blackmailing, identity theft, and other forms of cybercrime \cite{xu-etal-2020-personal, Bhardwaj2024}. For instance, studies have found that common conversational agents, such as Siri, can be prompted to expose sensitive user information that was unconsciously shared in previous interactions through targeted questioning \cite{xu-etal-2020-personal}. These risks underscore the need for advanced security measures to protect personal data and ensure HDTs operate within the limits of the individual's consent and privacy preferences.\par
Third, the prospect of HDTs persisting beyond an individual's lifetime raises profound ethical questions. While some research suggests that providing some sort of (digital) continuation of presence can offer comfort (see \textit{"Continuing Bonds Theory"} in Section \ref{subsubsec:Discussion:VisionsAndFutureApplications}), others emphasize that it might also hinder the natural grieving process for some, posing challenges to achieving emotional closure. For instance, a study found that externalized continuing bonds, such as illusions or hallucinations of the deceased, can lead to negative conflicts and impede the acknowledgment of death, thereby complicating the grieving process \cite{MartinezEsquivel2023}. Additionally, the enduring digital existence raises complex discussions over data ownership, rights to digital content, and the management of HDTs posthumously, highlighting the need for comprehensive guidelines to address these concerns.\par
Together, these challenges underscore how personal HDTs could profoundly benefit society but also pose significant ethical dilemmas, depending on their conceptualization and implementation. Therefore, to harness the benefits of HDTs while mitigating associated risks, a multi-disciplinary effort from technologists, ethicists, policymakers, and society is necessary to further explore the technological feasibility and potential applications while simultaneously developing robust frameworks that ensure the reliability, privacy, safety, and ethical integrity of HDTs. 

\section{Conclusion}
This paper introduced a novel HDT system that extends beyond traditional LLMs and RAG architectures by incorporating personalized memory retrieval, adaptive learning, and neural plasticity-inspired consolidation. By dynamically integrating dialogues, physiological data, and reflective insights, the system achieves a more authentic representation of individuals, enabling personalized and contextually aware interactions. Our results demonstrate the system’s ability to replicate an individual’s unique conversational style depending on who they are speaking with, while also enriching responses with dynamically captured personal experiences, opinions, and memories.\par

This approach unlocks a range of innovative applications, including autonomous HDTs that digitally represent individuals in professional and social contexts, a new paradigm for social media where users interact with HDTs of celebrities and friends, and virtual preservation of deceased loved ones to conserve knowledge and support the grieving process. However, the increasing sophistication of HDTs also raises ethical and security challenges, particularly concerning data privacy, identity management, and long-term societal implications. \par

We strongly encourage researchers to build upon our approach and explore both similar and alternative conceptualizations of HDTs, their potential applications, and the necessary advancements in autonomy, emotional intelligence, and ethical safeguards to ensure that HDTs remain trustworthy, secure, and aligned with human values.

\section*{Acknowledgments}  
The authors acknowledge the use of ChatGPT, an AI language model developed by OpenAI, in the preparation of this paper. ChatGPT was used to assist with content suggestions, refinement of writing, and improvements in clarity, coherence, and structure. However, all conceptual contributions, research findings, and final editorial decisions were made by the authors. \par

This project has received funding from the European Union’s Horizon 2020 research and innovation program under grant agreement No 871767.

%Bibliography
\bibliographystyle{unsrt}  
\bibliography{references}

\end{document}